\documentstyle[epsfig]{elsart}
\newcommand{ \be }{\begin{equation}}
\newcommand{ \ee }{\end{equation}}
\newcommand{ \bea }{\begin{eqnarray}}
\newcommand{ \eea }{\end{eqnarray}}

\newcommand{ \mpx }{\langle p_{x} \rangle}

\begin{document}

\begin{frontmatter}

\title{Directed flow of antiprotons
in Au+Au collisions at AGS}

\author[mcgill] {J.~Barrette},
\author[wayne]  {R.~Bellwied},
\author[wayne]  {S.~Bennett},
\author[stony]  {R.~Bersch},
\author[gsi]    {P.~Braun-Munzinger},
\author[stony]  {W.~C.~Chang}, 
\author[pitts]  {W.~E.~Cleland}, 
\author[pitts]  {M.~Clemen},
\author[idaho]  {J.~Cole}, 
\author[wayne]  {T.~M.~Cormier},
\author[mcgill] {Y.~Dai}, 
\author[bnl]    {G.~David}, 
\author[stony]  {J.~Dee}, 
\author[spaulo] {O.~Dietzsch}, 
\author[idaho]  {M.~Drigert},
\author[heidel] {S.~Esumi}, 
\author[heidel] {K.~Filimonov}, 
\author[wayne]  {A.~French}, 
\author[stony]  {S.~C.~Johnson},
\author[wayne]  {J.~R.~Hall}, 
\author[stony]  {T.~K.~Hemmick}, 
\author[heidel] {N.~Herrmann}, 
\author[gsi]    {B.~Hong},
\author[stony]  {Y.~Kwon},
\author[mcgill] {R.~Lacasse}, 
\author[wayne]  {Q.~Li}, 
\author[bnl]    {T.~W.~Ludlam},
\author[mcgill] {S.~K.~Mark},
\author[bnl]    {S.~McCorkle},
\author[gsi]    {D.~Mi\'{s}kowiec},
\author[bnl]    {E.~O'Brien},
\author[stony]  {S.~Panitkin},
\author[stony]  {V.~Pantuev},
\author[stony]  {P.~Paul},
\author[stony]  {T.~Piazza},
\author[stony]  {M.~Pollack},
\author[wayne]  {C.~Pruneau}, 
\author[mcgill] {Y. J. Qi},
\author[idaho]  {E.~Reber}, 
\author[mcgill] {M.~Rosati},
\author[stony]  {S.~Sedykh}, 
\author[wayne]  {J.~Sheen}, 
\author[pitts]  {U.~Sonnadara}, 
\author[heidel] {J.~Stachel},
\author[mcgill] {N.~Starinski},
\author[spaulo] {E.~M.~Takagui},  
\author[mcgill] {V.~Topor~Pop}, 
\author[stony]  {M.~Trzaska},
\author[wayne]  {S.~Voloshin},
\author[stony]  {T.~B.~Vongpaseuth},
\author[mcgill] {G.~Wang}, 
\author[heidel] {J.~P.~Wessels}, 
\author[bnl]    {C.~L.~Woody}, 
\author[stony]  {N.~Xu},
\author[stony]  {Y.~Zhang} and
\author[stony]  {C.~Zou}\\ 
(E877 Collaboration)

\address[bnl]{Brookhaven National Laboratory, Upton, New York 11973}
\address[gsi]{Gesellschaft f\"ur Schwerionenforschung, 64291 Darmstadt,
  Germany}
\address[heidel]{Physikalishes Institut, Universit\"at Heidelberg, 69120
  Heidelberg, Germany}
\address[idaho]{Idaho National Engineering Laboratory, Idaho Falls,
  Idaho 83402}
\address[mcgill]{Department of Physics, McGill University, Montr\'eal,
  Qu\'ebec, Canada H3A 2T8}
\address[pitts]{Department of Physics and Astronomy, University of
  Pittsburgh, Pittsburgh, Pennsylvania 15260}
\address[stony]{Department of Physics and Astronomy, State University of
  New York at Stony Brook, New York 11794}
\address[spaulo]{Department of Physics, University of S\~ao Paulo, 05508 S\~ao
  Paulo, Brazil}
\address[wayne]{Physics Department, Wayne State University, Detroit,
  Michigan 48202}

\begin{abstract}
Directed flow of antiprotons is studied in
Au+Au collisions at a beam momentum of 11.5A GeV/c. 
It is shown that antiproton directed flow 
is anti-correlated to proton flow.
The measured transverse momentum dependence of the antiproton flow is
compared with predictions of the  
RQMD event generator.\\

\end{abstract}

\end{frontmatter}

\section{Introduction}

The production of antiprotons at AGS energies (10-15A GeV) is near 
threshold in nucleon-nucleon collisions. Via collective effects,
$\bar{p}$ production can be enhanced in nucleus-nucleus collisions as compared 
to p+p collisions \cite{mik90,koch89,koch91,schaf91}.
The observed yield of antiprotons 
is a result of both production and subsequent 
annihilation.
The high antiquark densities in the quark-gluon plasma (QGP) state might
lead to antibaryon abundances greatly exceeding the values in
a chemically equilibrated hadron gas \cite{heinz86}. However, due to
conservation of entropy, those abundances may be considerably diluted to
a level very close to chemical equilibrium \cite{heinz89}.
On the other hand, antibaryons have a large annihilation 
cross section, which  may be modified in the baryon rich colliding systems
\cite{kahana93}.
Measurements of $\bar{p}$  at the AGS may also contain a large feed-down
contribution from the decay of antilambdas, 
$\bar{\Lambda}\;\rightarrow\;\bar{p} + \pi^+ $, as well as from other 
antihyperons ($\bar{Y}$).
Microscopic calculations have given some indication of how the annihilation 
process might change in the dense environment of a heavy-ion 
collision. It has been shown that the amount of annihilation in the
nuclear medium depends strongly on the formation time of the hadrons 
\cite{kumar94}. In view of all this, it came as a surprise that both 
in Si+Au and Au+Au collisions the measured $\bar{p}/p$ ratio is well 
described by computing the yields from a fireball in complete thermal 
and chemical equilibrium \cite{pbm95,johanna96}. Since the 
conditions for thermal equilibrium are equal to those for 
hydrodynamic evolution of the fireball, it is of great interest to 
determine experimentally the flow features for antiprotons. 

The hydrodynamic evolution of the fireball leads to characteristic 
flow patterns such as radial expansion \cite{pbm95} and azimuthal 
anisotropies, first discovered in ultra-relativistic nuclear collisions 
by ref. \cite{l877flow1}. Since then, the study of anisotropies in the 
azimuthal distribution of particles, also called
anisotropic (directed, elliptic, etc.) transverse flow, plays an important
role in high energy nuclear collisions 
\cite{lqm97,sorge99,cassing99}. Detailed measurements (for surveys see refs. \cite{ritt98,herr99}) have revealed that 
in a non-central high energy heavy-ion collision, nucleons in the backward hemisphere are preferentially 
emitted in the direction of the impact parameter vector pointing from
the target to the projectile. Antiprotons co-moving 
with those nucleons have a greater probability of 
annihilation and rescattering, and this could result in an 
anticorrelation with the nucleon directed flow - the so-called 
{\em anti-flow} of antiprotons in nuclear collisions 
predicted by Jahns {\it et al.} \cite{jahns94a,jahns94b}.
Measurements of the directed flow of antiprotons are important for
understanding the role of annihilation in dense nuclear matter.
They could also provide insight into the 
mechanism of anti-quark production in heavy-ion collisions. 
A surprising result reported in refs. \cite{arms99,back99}, is the possible anomalous enhancement of the 
$\bar{\Lambda}/\bar{p}$ ratio even beyond values expected for complete chemical equilibration \cite{pbm95}. 
One possible explanation could be that  $\bar{\Lambda}$ baryons
have a lower annihilation cross section than $\bar{p}$.  
A straightforward way to test this hypothesis
would be the measurement of $\bar{p}$  and $\bar{\Lambda}$
{\em (anti-)flow} \cite{miklos99}. 
 
Antiproton production in heavy-ion induced reactions at the AGS
has been studied experimentally by the  
E858/E878 \cite{aoki92,benet97}, E814/E877 \cite{baret93,green92}, 
E866 \cite{sako96}, E886 \cite{dieb93}, E864 
\cite{rotondo96,arms99}, E802 \cite{abbott93,ahle98}, and E917 \cite{back99} collaborations.
In the current paper we present the first results on the 
measurements of directed flow of antiprotons
detected in the E877 spectrometer in Au+Au 
collisions at a beam momentum of 11.5A GeV/c.
The results were obtained from the analysis of 
46 millions of central Au+Au events recorded during the 1995 heavy-ion
 run. 

\section{E877 apparatus}

\begin{figure}[hbt!]
\centering
 \epsfig{file=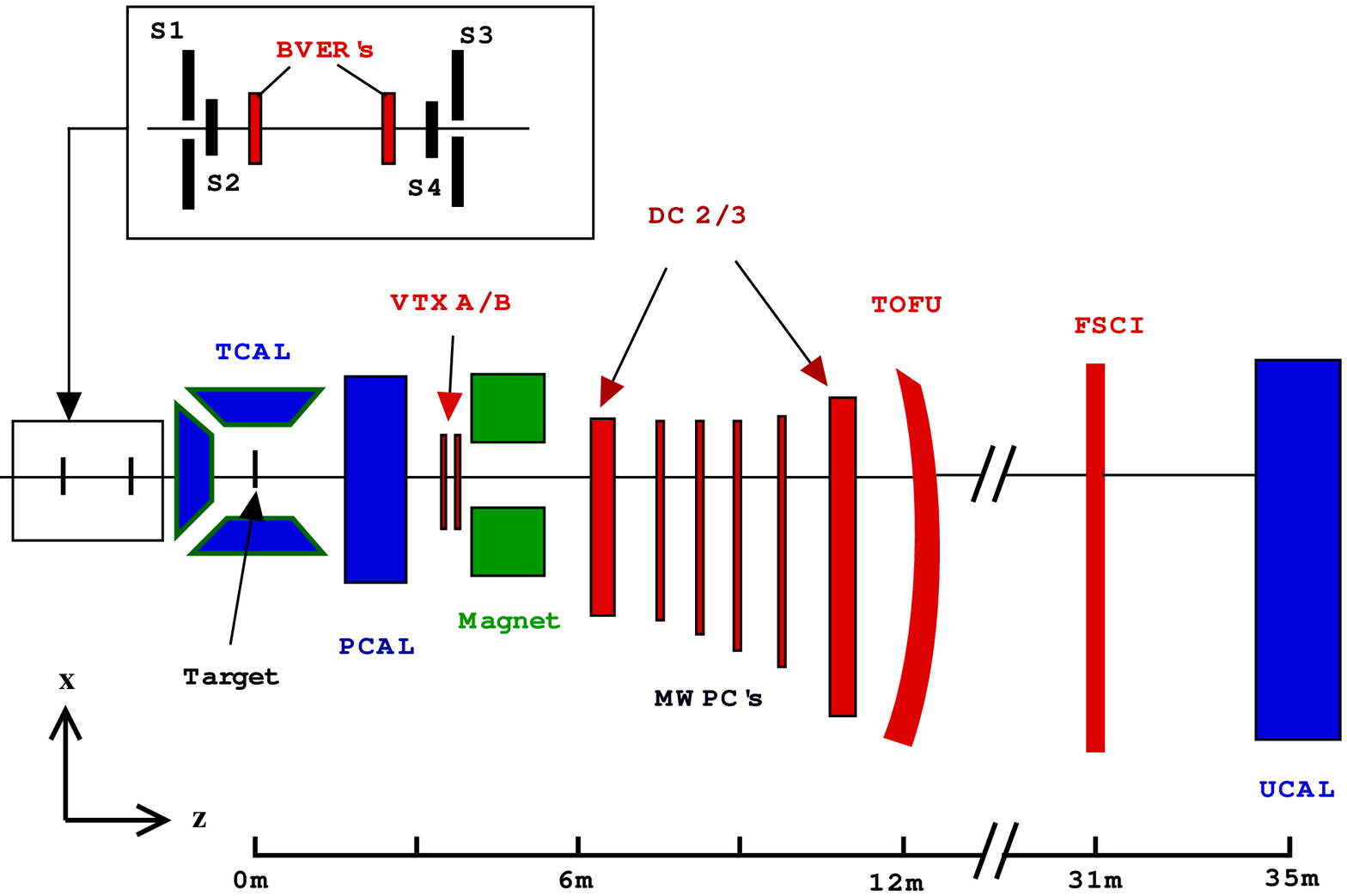,width=14cm, clip=}
 \vskip 0.5cm
\caption{E877 experimental setup.}
 \label{fig:setup_all}

\end{figure}

Fig.~\ref{fig:setup_all} shows the E877
experimental setup which is also discussed in
\cite{l877flow1,l877flow2,l877flow3}. 
The trajectory of each beam particle was measured by two silicon
micro-strip beam vertex
detectors (BVER), located
at 2.8 m and 5.8 m upstream of the target. 
For the 1995 run, each detector was upgraded from single-sided
silicon wafers with one-dimensional pitch of 50 $\mu$m to  
double-sided wafers with a 200~$\mu$m pitch in both the 
$x$ and $y$ directions \cite{ydai}. Using these detectors 
the coordinates and angle of beam particles at the target were determined 
with an accuracy of 300 $\mu$m in position and 60 $\mu$rad in angle. 

The determination of the centrality of the collision and of the reaction
plane orientation were done using the transverse energy distribution measured in
the target calorimeter (TCAL), and participant calorimeter (PCAL).  Both
calorimeters had $2 \pi$ azimuthal coverage and, combined,
provided nearly complete polar angle coverage: TCAL and PCAL covered the
pseudorapidity regions $-0.5 <\eta <0.8$ and $0.8 <\eta < 4.2$,
respectively~\cite{l877flow2}. 

Charged
particles emitted in the forward direction and passing through a
collimator ($ -134$ mrad$ < \theta_{horizontal} < 16 $ mrad, $ -11$
mrad$ < \theta_{vertical} < 11 $ mrad) were analyzed by a high resolution
magnetic spectrometer with a horizontal bend-plane.  The spectrometer acceptance covered mostly the
forward rapidity region.  The momentum of each particle was measured
using two drift chambers, DC2 and DC3, 
whose pattern recognition capability was aided by four
multi-wire proportional chambers (MWPC).  The average momentum
resolution was $\Delta p/p \approx$ 3\%, limited by multiple scattering.  A
time-of-flight hodoscope (TOFU) located directly behind the tracking
chambers provided the time-of-flight information with an average resolution of
85~ps.  Energy loss measurements in TOFU and in a Forward Scintillator (FSCI) array located
approximately 30~m downstream of the target were used to determine the
particle charge.

For the 1995 run, the spectrometer was upgraded with upstream
tracking chambers. Two identical multi-wire
proportional chambers with highly
segmented chevron-shaped readout pads, VTXA
and VTXB, were instrumented and placed at 2 m and 2.25 m downstream of the target, 
just in front of
the spectrometer magnet. They provided a precise measurement (about
$300~\mu$m resolution) of the track coordinate 
in the bending plane of the spectrometer
before deflection in the
magnetic field.
 A description of the design, implementation and performance of 
the vertex detectors can be found in \cite{bersch95}.

\section{Antiproton identification}
\begin{figure}[hbt!]
\centering
 \epsfig{file=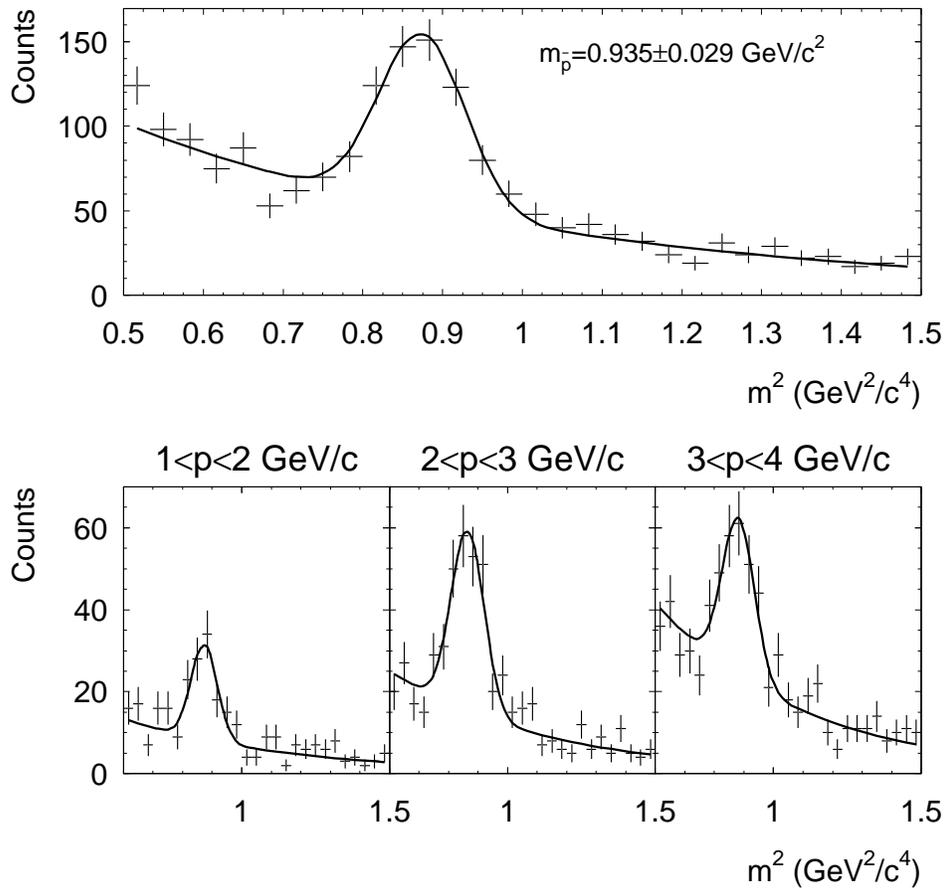,width=14cm,clip=}
 \vskip 0.5cm
\caption{Upper plot: the mass squared distribution in the 
antiproton peak region for negative
 particles with momenta less than 4 GeV/c. 
The Lower plot shows the same distribution 
for three different momentum intervals. 
The solid line is a fit by a Gaussian and  
an exponential.}
\label{fig:pbar_mass}
\end{figure}

The particle identification was performed by combining measurements of
momentum, velocity, and charge of the particle. 
To reduce the
background in the identification of antiprotons, tracks were required
to have a confirmation from both upstream
VTX detectors. Tracks were also required to have complete
information from the cathode pad readout in both drift chambers.
Fig.~\ref{fig:pbar_mass} shows the distribution of particle mass
squared,
calculated from the measured momentum and velocity, for negative
particles with momentum less than 4 GeV/c. A clear antiproton peak is
seen, with good signal-to-background
ratio.
The contribution
of the background increases with momentum (see lower panel of
Fig.~\ref{fig:pbar_mass}) and becomes dominant at
$p>4.5$ GeV/c. A maximum momentum of 4 GeV/c was required for clean 
identification of antiprotons. 
The mass resolution is quantitatively understood in terms of the
intrinsic detector resolution, multiple scattering and time-of-flight
resolution. For a positive antiproton identification we selected, in a plot of momentum versus the measured mass, 
a region of $\pm 2.0 \sigma_{m^2}(p)$ around the antiproton mass peak, where
$\sigma_{m^2}(p)$ represents the resolution in the mass squared 
at a given particle momentum $p$. The dependence of 
the mass resolution on momentum was assumed 
to be equal to that of protons. 

\begin{figure}[hbt!]
\centering
 \epsfig{file=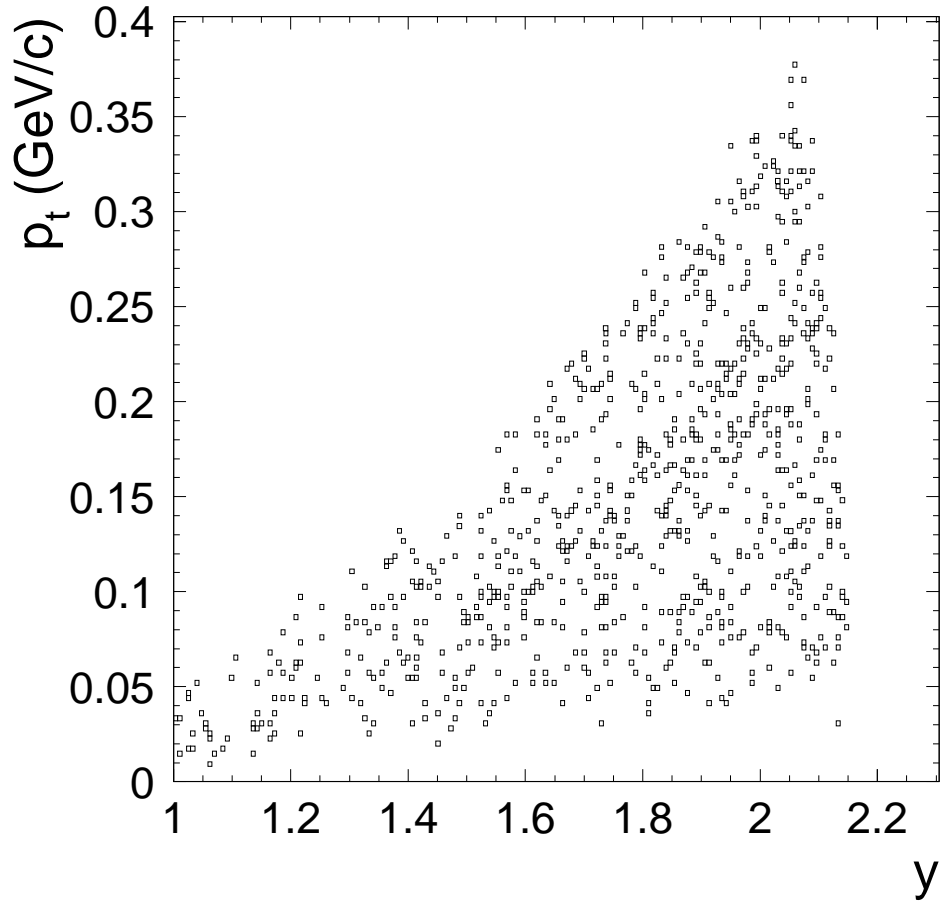,width=14cm, clip=}
 \vskip 0.5cm
\caption{Antiproton acceptance in the ($p_t,y$) phase space covered by the E877
spectrometer. Midrapidity is $y=1.6$.}
 \label{fig:pbar_acc}
\end{figure}

Fig.~\ref{fig:pbar_acc} shows the antiproton acceptance in the
transverse momentum $p_t$ and rapidity $y$  
coordinates.
The E877 spectrometer mainly covers the low $p_t$ region
($p_t<0.4$ GeV/c) forward of $y_{cm}=1.6$ ($1.6<y<2.2$).
The final data sample comprises about 750 antiprotons with
about 400 in the $1.8<y<2.2$ rapidity region.

Since tracking only starts 2 m downstream of the target, the E877
spectrometer is not well suited for separation of primary antiprotons from
those fed down from anti-hyperon decays. Because of the large mass
asymmetry and small decay momentum in these decays, the antiproton track 
points back to the target. A
previous extensive study \cite{kwon97}
based on a Monte-Carlo simulation showed that antiprotons
are reconstructed for 70\% of the $\bar{\Lambda}$ and for 100\% of the
$\Sigma^-$ decaying such that the $\bar{p}$ is emitted into the
acceptance of the spectrometer and consistent with originating from the target.

\section{Results and Discussion}
\label{antip_dis}

\subsection{Experimental data}

The azimuthal anisotropy in particle production is studied by means of
Fourier analysis of azimuthal
distributions \cite{l877flow1,l877flow2,l877flow3,lvzh,lolli} with
respect to the reaction plane. We
study the rapidity, transverse momentum, and centrality dependence of
the Fourier coefficients $v'_n$ (amplitude of $n$-th harmonic) in the
decomposition:
\[
E \frac{d^3 N}{d^3 p} =  
\frac{1}{2\pi} \frac {d^2N}{p_t d p_t dy}(1+2 v'_1 \cos \phi
+2 v'_2 \cos 2 \phi
+...),\]
where the azimuthal angle $\phi=\phi_{lab}-\psi_r$ is taken with respect to the
 reaction plane orientation. Directed flow is quantified by the dipole
 Fourier coefficient $v'_1$. 

Similarly to the analysis performed in \cite{l877flow2,l877flow3}, the
reaction plane angle is determined from the measured transverse energy
distribution in four non-overlapping
pseudorapidity windows. The {\em reaction plane resolution}, i.e. the
accuracy with which the reaction plane orientation is determined, is
evaluated by studying the correlation between flow angles measured in
different windows.  Finally, the flow signals are corrected for the
finite reaction plane resolution.  Details of this procedure are
described in \cite{l877flow2,l877flow3}. In the following, only
coefficients $v_1$ corrected for the reaction plane resolution will be shown.

\begin{figure}[hbt!]
\centering
 \epsfig{file=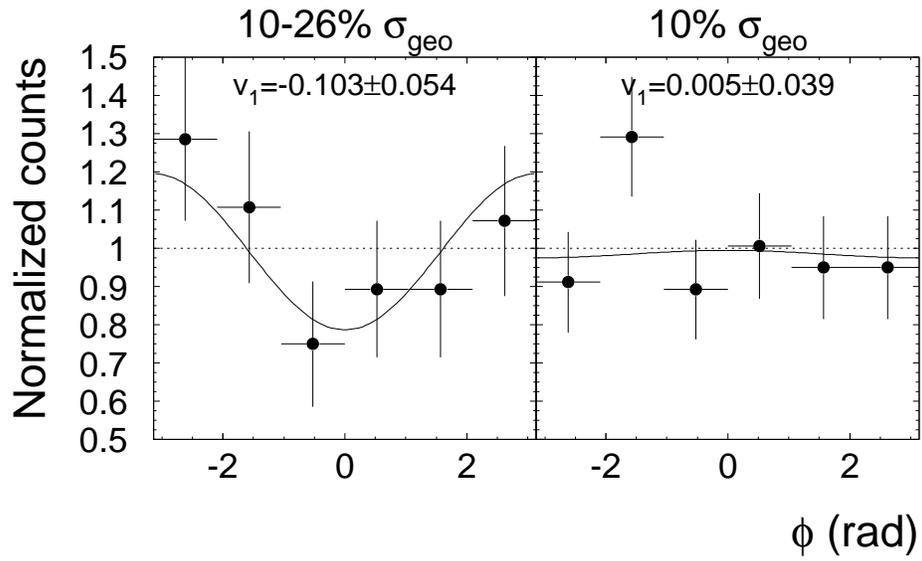,width=14cm, clip=}
\vskip 0.5cm
\caption{Antiproton azimuthal distributions measured for rapidity
 $1.8<y<2.2$ and two collision centralities. The distributions are
 normalized such that the average is unity. The solid line is a fit by 
a $1+2v_1\cos\phi$ distribution.}
 \label{fig:pbar_phi}
\end{figure}

The experimental acceptance for $\bar{p}$ is close to mid-rapidity
(Fig.~\ref{fig:pbar_acc}), where the directed flow changes sign. To
optimize a possible flow signal, we skip the region very close to
mid-rapidity and present, in Fig.~\ref{fig:pbar_phi}, 
the measured azimuthal distributions of antiprotons emitted
at rapidity $y>1.8$, for two collision centralities. 
To determine $v_1$, the distributions were fitted with a distribution $f(\phi)=1+2v_1\cos\phi$.
A pronounced minimum is observed at $\phi=0$
for semicentral collisions ($\sigma = 10-26\%~\sigma_{geo}$),
indicating that the antiproton production is strongly anti-correlated with
the flow of nucleons. For more central collisions (top 10\% of $\sigma_{geo}$), 
the antiproton azimuthal
distribution is, within errors, isotropic ($v_1\approx 0$).
No or very little proton flow has been observed for the most
central Au+Au collisions \cite{l877flow3}.

In order to check the contribution of the background to the observed flow
signal we evaluated the azimuthal distributions for the particles whose
measured mass was outside the window for positive antiproton
identification. For the background sample we selected, in a plot of
momentum versus the measured mass, a region of $\pm$2-6 $\sigma_{m^2}(p)$ away
from the $\bar{p}$ mass peak. On the low mass side we cut the region less
than 3.5 $\sigma_{m^2}(p)$ away from the $K^-$ peak.  The azimuthal
asymmetry for the background sample was found to be consistent with zero
($v_1=-0.023\pm0.053$ for 10-26\% $\sigma_{geo}$ centrality window and
$v_1=0.021\pm0.038$ for 0-10\% $\sigma_{geo}$ centrality window). We also
checked the $v_1(p_t)$-dependence for the background sample and found no
signal within the statistical errors for all $p_t$ bins. 
The data shown do not include the correction due to background subtraction.
Assuming a typical signal-to-background ratio 
of 1.2 and that background has no asymmetry, the average $v_1$ for 10-26\% $\sigma_{geo}$ centrality window increases to $\approx-0.19\pm0.07$. 

The measured transverse momentum dependence of $v_1$ is shown in Fig.~\ref{fig:pbar_v1}.
Large negative values of $v_1$ are observed for $p_t>0.1$ GeV/c. It
should be noted that the absolute value of the proton flow signal at 
similar rapidities and
transverse momenta is significantly smaller \cite{l877flow3}. The right
hand panel shows a result consistent with zero flow for the more central
bin.

\subsection{RQMD predictions}

The Relativistic Quantum Molecular Dynamics (RQMD) model
\cite{Sorge:1989vt}
has been widely used in describing relativistic heavy-ion collisions. 
It combines 
classical propagation of all hadrons with string and resonance
excitations in the primary collisions of nucleons from 
target and projectile. Overlapping color strings may fuse into 
so-called ropes \cite{sorge95}. 
Subsequently, the fragmentation products from rope,
string and resonance decays interact with each other and 
with the original nucleons, mostly via binary collisions. 
These interactions drive the system towards equilibration \cite{sorge96}
and are responsible for the development of collective flow, even in 
the pre-equilibrium stage. 
If baryons are surrounded by other baryons they acquire effective masses.
These effective masses are generated by introducing Lorentz-invariant 
Skyrme-type
quasi-potentials into the mass-shell constraints for the momenta,
which simulates the effect of a {\em mean field} \cite{sorge97}.
There are no potential-type interactions in the so-called {\em cascade mode} of
RQMD. 

For antiprotons, RQMD combines a large enhancement of initial production
via hadronic multi-step processes with strong absorption through the 
free $\bar{p}$ absorption cross section \cite{jahns92},
which results in the absorption of a large
fraction of the produced antiprotons ($\approx$ 90\% of $\bar{p}$ 
are reabsorbed for minimum bias Au+Au collisions \cite{jahns94b}). 
The newest version 
of the model (RQMD v2.3) uses the
formation of a quasi-bound state (proton-antiproton
molecule) to significantly reduce the possibility of antiproton
annihilation. 
However, the presence of a 
mean color potential could possibly have a much more 
significant effect on the final antiproton distribution \cite{spieles96}.
A systematic study of the antibaryon production 
with various projectile-target combinations and at 
different energies 
would be useful for refining this approach and for
understanding the annihilation process in relativistic heavy-ion 
collisions. 

\begin{figure}[hbt!]
\centering
 \epsfig{file=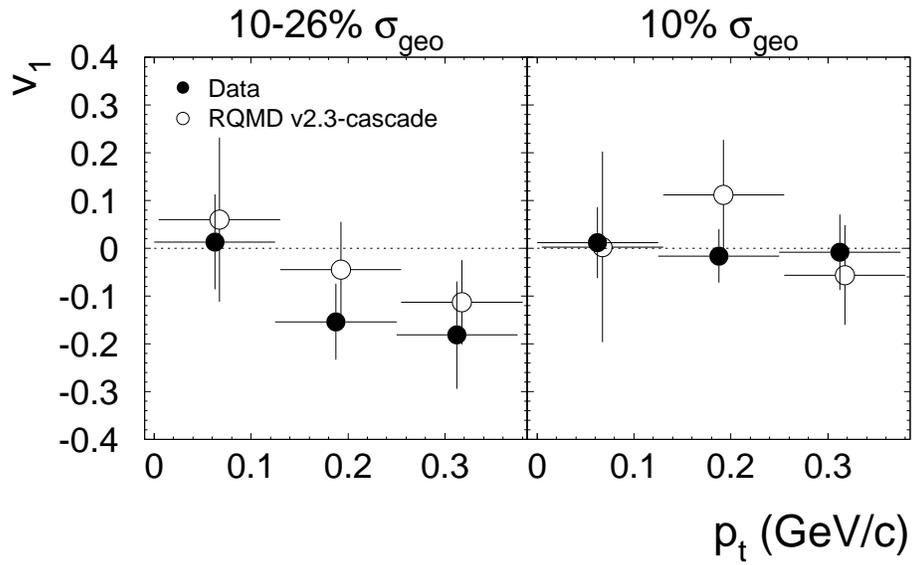,width=14cm, clip=}
\vskip 0.5cm
\caption{$v_1(p_t)$ of antiprotons for rapidities 
 $1.8<y<2.2$ and two collision centralities. The data (solid circles) are
 compared with the calculations using the RQMD v2.3 - {\em cascade mode}
 (open circles).}
 \label{fig:pbar_v1}
\end{figure}

Predictions of the RQMD model 
run in {\em cascade mode} for the $p_t$-dependence of $v_1$ for antiprotons
are compared to
the data in Fig.~\ref{fig:pbar_v1}.
For the model, we only used antiprotons produced in
the experimental acceptance window.
The magnitude of the negative signal observed in the
data for semicentral collisions is comparable, within the errors, to
that predicted by the model although, systematically, the model values
are somewhat smaller. 

\begin{figure}[hbt!]
\centering
 \epsfig{file=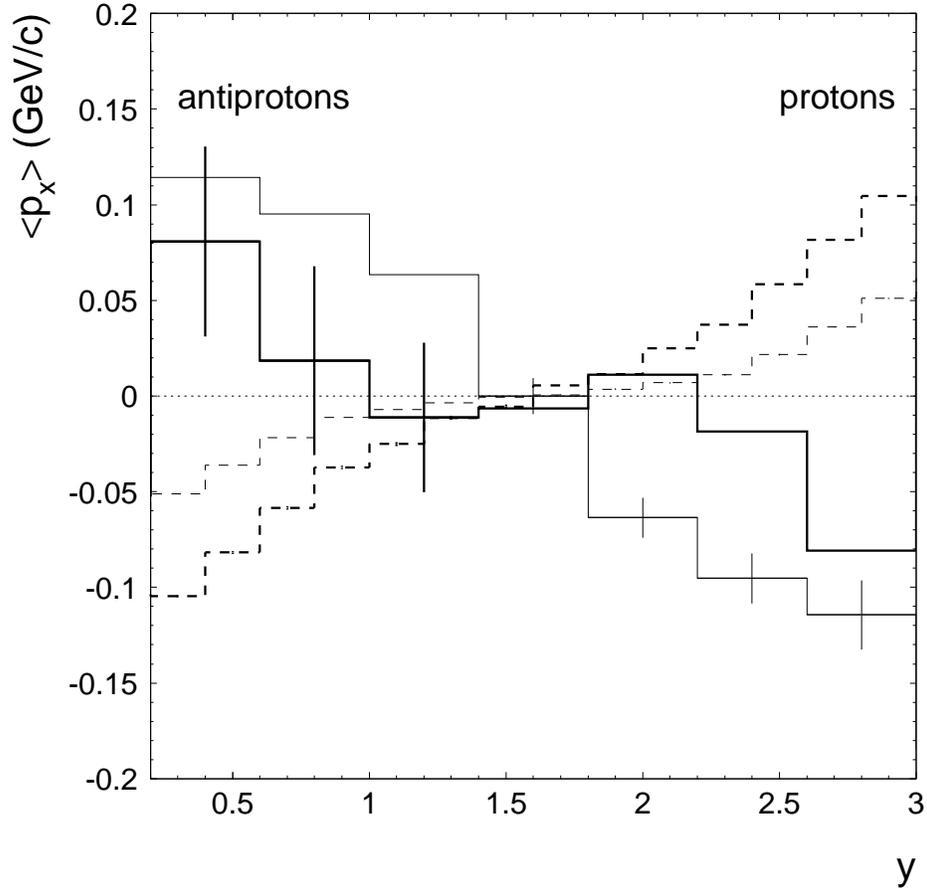,width=14cm, clip=}
\vskip 0.5cm
\caption{Mean values of the transverse momentum projected into the
  reaction plane $\mpx$ of protons (dashed lines) and antiprotons (solid
  lines)
  as a function of rapidity for Au+Au collisions at 11.5A GeV/c 
with impact parameters
$b<10$ fm generated by using the RQMD v2.3 ({\em cascade
  mode} - thin lines; {\em mean field mode} - thick lines).}
 \label{fig:rqmd_mpx}
\end{figure}

A current question of interest and debate is the relative importance of
including nuclear mean field effects in intra-nuclear cascade models.  At
lower energies, experimental data and a detailed comparison with
theoretical predictions have been discussed in \cite{cassing99}. 
An attractive $\bar{p}$ potential of the order of 100 to 150 MeV at normal 
nuclear density is needed to reproduce the size and shape of the 
experimental spectra \cite{sibir98}.
It is expected that collective flow may be an 
important observable for evaluating the importance of mean field effects
\cite{spieles96}. 
It was shown, in particular that,
in semicentral Au+Au collisions, protons exhibit a strong azimuthal
anisotropy, with the amplitude of flow well reproduced by RQMD only if
the effects of {\em mean field} are included \cite{l877flow3}. 
It should be noted, however that, even including the mean field,
only the $p_t$-integrated signal is reproduced by the model, 
while the predicted $p_t$-dependence significantly differs from the data.
In Fig.~\ref{fig:rqmd_mpx} 
we present the RQMD v2.3 calculations for proton and antiproton mean
values of the transverse momentum projected onto the reaction plane,
$\mpx$, as a function of rapidity for Au+Au collisions at 11.5A GeV/c 
with impact parameters $b<10$ fm. 
Calculations are presented for a pure cascade and mean-field modes of RQMD.
Since the calculations are very
CPU intensive we have a
limited antiproton statistics for the mean field calculations: about 500
$\bar{p}$'s compared to more than 8700 $\bar{p}$'s for the cascade mode.
The flow of nucleons is enhanced due to additional repulsion
in the mean field approach. For antiprotons, on the other hand, 
the effective mean field potential 
in nuclear matter is expected to be attractive.
The antinucleons 
are pulled towards the nucleon source, therefore the {\em anti-flow}
should be considerably weakened \cite{sorge_com}, as observed
in Fig.~\ref{fig:rqmd_mpx}.
Such a weakening is also supported by the results of Spieles {\it et al.} \cite{spieles96}, where it was shown that,
when the influence of the real
part of an antinucleon-nucleus optical potential 
on the $\bar{p}$ momentum spectra is included into the calculations,
the antiproton annihilation is indeed suppressed in the heavy-ion 
environment, such that the sum of in-medium
annihilation cross section and elastic cross section 
is closer to the free annihilation cross sections which are 
employed in the cascade mode of RQMD v2.3.
On the other hand, the amplitude of the measured antiproton flow signal
is rather well described by the pure cascade
calculations (see Fig.~\ref{fig:pbar_v1}). A similar observation has
been recently made by comparing the measurements of $K^+$ directed
flow with the cascade and mean field RQMD predictions \cite{barqm99},
where it has been shown that the cascade calculations reproduce the data
better than the mean field calculations. 
Further measurements 
and detailed calculations in the {\em mean field mode} 
are required and this question cannot be fully settled until
antiproton yield and flow have been measured over a larger phase space. 

\section{Conclusion}
 
 We have observed, for the first time, a strong azimuthal anisotropy in
 antiproton production in semicentral Au+Au collisions which is anti-correlated
with the nucleon emission.
The observed large {\em anti-flow} for $\bar{p}$ indicates that 
a strong annihilation process is involved in the dense 
nuclear matter as predicted by previous theoretical studies.
The presented data have confirmed the need for a thorough
 investigation, both experimentally and theoretically, of antibaryon
 production in nucleus-nucleus 
interactions. Whereas the measured antiproton yields/spectra can be described 
within a thermal model assuming a chemical/thermal equilibration and 
collective expansion of the system, antiprotons exhibit a directed 
flow which is even stronger 
than expected by the models incorporating mean-field effects.
 A complementary measurement of the directed flow of antihyperons such as
$\bar{\Lambda}$ in the future combined with our results
would provide necessary information for the
decoupling of the flow contribution of primary antiprotons and that of
antihyperons. Furthermore, the comparison between the directed flow
behavior of these two contributions could give crucial information for
understanding the observed unexpected $\bar{\Lambda}/\bar{p}$
ratios.
 
\section*{Acknowledgments}

We thank the AGS staff, W. McGahern and Dr. H. Brown for excellent
support and acknowledge the help of R. Hutter in all
technical matters. Financial support from the US DoE, the NSF, the
Canadian NSERC, and CNPq Brazil is gratefully acknowledged.


\newpage 

\newpage 

\newpage

\newpage 

\newpage 

\newpage

\end{document}